\documentclass[twocolumn,superscriptaddress,pra]{revtex4-1} 
\usepackage{color}
\usepackage{xspace}
\usepackage{amsmath}
\usepackage{amssymb}
\usepackage{amsbsy}
\usepackage{graphicx}
 \usepackage{bm}
 \usepackage{float}
\usepackage{relsize}

\newcommand{\x}{\mathbf{x}}
\newcommand{\br}{\boldsymbol\rho}
\newcommand{\brp}{\boldsymbol\rho^\prime}

\newcommand{\bk}{\mathbf{k}_\rho}

\begin{document}

\title{Enhanced quantum spin fluctuations in a binary Bose-Einstein condensate}

\author{R.~N.~Bisset}
\email{rnbisset@gmail.com}
\affiliation{INO--CNR BEC Center and Dipartimento di Fisica, Universit\`a di Trento, 38123 Povo, Italy}

\author{P.~G.~Kevrekidis}
\affiliation{Department of Mathematics and Statistics, University of Massachusetts,
Amherst, Massachusetts 01003-4515 USA}

\author{C.~Ticknor}
\affiliation{Theoretical Division, Los Alamos National Laboratory, Los Alamos, New Mexico 87545, USA}

\begin{abstract}

For quantum fluids, the role of quantum fluctuations may be significant in several regimes such as when the dimensionality is low, the density is high, the interactions are strong, or for low particle numbers.
In this paper we propose a fundamentally different regime for enhanced quantum fluctuations without being restricted by any of the above conditions.
Instead, our scheme relies on the engineering of an effective attractive interaction in a dilute, two-component Bose-Einstein condensate (BEC) consisting of thousands of atoms.
In such a regime, the quantum spin fluctuations are significantly enhanced (atom bunching with respect to the noninteracting limit) since they act to reduce the interaction energy - a remarkable property given that spin fluctuations are normally suppressed (anti-bunching) at zero temperature.
In contrast to the case of true attractive interactions, our approach is not vulnerable to BEC collapse.
We numerically demonstrate that these quantum fluctuations are experimentally accessible by either spin or single-component Bragg spectroscopy, offering a useful platform on which to test beyond-mean-field theories.
We also develop a variational model and use it to analytically predict the shift of the immiscibility critical point, finding good agreement with our numerics.

\end{abstract}

\maketitle


\section{Introduction}

Quantum fluctuations are ubiquitous in nature, lying at the heart of a wide variety of physical phenomena ranging from, for example, the Van der Waals force and the Casimir effect, through to Hawking radiation of black holes and the theory of cosmic inflation, providing the initial seed for the large-scale structure of the universe today. Despite their importance, quantum fluctuations are notoriously difficult to describe theoretically and many important questions remain unanswered \cite{SachdevBook,Pitaevskii16}.

Focusing on quantum fluids with short-range interactions, the role of quantum fluctuations becomes important in several qualitatively distinct regimes.
Among the most dramatic of these are the one-dimensional systems, for which quantum fluctuations are so large that long-range order is destroyed \cite{Mermin1966,Hohenberg1967}.
Another well-known regime is that of high density (and strong interactions) where, for example, in the superfluid phase of liquid $^{4}$He, quantum fluctuations cause a depletion of the underlying Bose-Einstein condensate (BEC) of around 90\% \cite{Dalfovo1999}.

Dilute quantum gases have accelerated quantum-fluctuation research in recent years, offering a clean and highly-controllable testbed in which nearly all aspects are tunable, such as the interaction strength, confinement, mass and particle number \cite{Dalfovo1999,Leggett2001}.
This has allowed a series of innovative experiments to probe the emergence of quantum fluctuations by approaching the high-density regime from the dilute limit \cite{Shin2008,Lopes2017,Navon2010,Navon2011,Papp2008A,Armijo2012,Altmeyer2007,Pollack2009,Chang2016}.
More recently, experimental groups were able to create dilute, self-bound droplets with liquid properties using dipolar \cite{Schmitt2016,Chomaz2016} and two-component condensates \cite{Cabrera2017,Semeghini2017}.
Remarkably, these are stabilized against collapse by quantum fluctuations thanks to an almost complete cancellation of the various mean-field contributions \cite{Petrov2015,Wachtler2016,Lima2011,Bisset2016,Saito2016}.

Quantum fluctuations can also be important for the regime of very small particle number $N$, typically when $N\sim10$.
Several groups have proposed using ultra-dilute quantum gases to investigate these in different ways, including: vortex nucleation in slowly rotating traps \cite{Parke2008,Dagnino2009}; the transition between Rabi-oscillation and self-trapped regimes in double well potentials  \cite{JuliaDiaz2010}; the generation of macroscopic state superpositions in rotating ring superlattices \cite{Nunnenkamp2008}; as well as few-boson systems in one dimension \cite{Garcia2014,Garcia2014A}.
Furthermore, the enhancement of quantum fluctuations near the superfluid to Mott-insulator phase transition occurs for small atom number per lattice site \cite{Greiner2002,Jaksch1998}.

In this paper, we propose an alternative regime for enhancing quantum fluctuations that does not require low dimensionality, high density, strong interactions or small $N$.
Our scheme instead relies on the engineering of effectively attractive interactions for the spin excitations of two-component (binary) BECs in the ground state, while the actual intracomponent and intercomponent contact interactions remain repulsive.
By {\it spin excitations}, we mean the out-of-phase excitations of the two components.
The interactions are effectively attractive in the sense that spin fluctuations act to decrease the interaction energy.
Contrary to the case of true attractive contact interactions \cite{Bradley1995,Donley2001}, the proposed regime is not vulnerable to collapse of the underlying BEC.

Numerous groups have experimentally realized binary BECs by combining different elements \cite{Thalhammer2008,McCarron2011,Lercher2011,Pasquiou2013,Wacker2015,Wang2016}, isotopes \cite{Papp2008,Sugawa2011,FerrierBarbut2014}, hyperfine states \cite{Myatt1997,Hall1998B,Matthews1999,Hoinka2012}, and even different spin states \cite{Maddaloni2000}. There has also been considerable theoretical attention, e.g.~see \cite{Ho1996,Pu1998,Ohberg1999,Trippenbach2000,Svidzinsky2003,Kasamatsu2004,Dutton2005,Mertes2007,Ronen2008,Oles2008,Anderson2009,Hofmann2014,Roy2015,Liu2016}.
In the thermodynamic (TD) limit, a binary condensate is miscible for small intercomponent scattering length $a_{12}$, up until $a_{12}^{\rm c,TD} \equiv \sqrt{a_{11}a_{22}}$, above which it phase separates and becomes immiscible \footnote{We have assumed that the particles of both components have the same mass}.
Interestingly, at finite temperature ($T$), long-wavelength spin fluctuations tend to diverge on approach to the immiscibility phase transition, while at $T=0$ they do not
\cite{Bisset2015,Bienaime2016}.
In fact, they instead approach the noninteracting, quantum shot-noise limit \cite{Astrakharchik2007,Klawunn2011}.
This occurs because at the transition $a_{12}^{\rm c,TD}$ the interactions have no preference for miscibility over immiscibility, and so the spin fluctuations behave as if the BEC was noninteracting.

We demonstrate that the situation changes dramatically, however, if one enters the regime in which quantum pressure becomes important. Here the critical point $a_{12}^{\rm c}$ is shifted higher than the thermodynamic-limit prediction, i.e. $a_{12}^{\rm c}>a_{12}^{\rm c,TD}$, thus opening a gap \cite{Navarro2009,Wen2012}.
For such a system, if the intercomponent scattering length lies within this gap, i.e.~$a_{12}^{\rm c,TD}<a_{12}<a_{12}^{\rm c}$, then quantum spin excitations experience an effective attractive interaction,
in the sense that they act to decrease the interaction energy.
We show that the resulting spin fluctuations are then greatly enhanced.
An intuitive picture for the physics at play: the trap pins the two components together, forcing miscibility of the ground state solution within the regime where the interactions would otherwise prefer immiscibility.
Quantum spin fluctuations are then enhanced as a reward for lowering the interaction energy.

By numerically solving the Gross-Pitaevskii equations (GPE) and Bogoliubov de Gennes (BdG) equations, we demonstrate that such quantum spin fluctuations should be experimentally observable via either spin or single-component Bragg spectroscopy. The advantage of Bragg spectroscopy is that thermal contributions cancel, to first order  \cite{Brunello2001,Blakie2002}, resulting in the observation of the quantum fluctuations.
Moreover, we develop a variational model and use this to analytically predict the shift of the critical point, providing a simple means for finding suitable regimes of enhanced quantum spin fluctuations. We find good agreement between the predictions of the variational model and the numerical results.

.

\section{Formalism}

\subsection{System and Parameters}\label{Sec:System}

We investigate a binary condensate for which interspecies interconversion is prohibited and the populations are fixed.
The trapping potential is harmonic $V(\x) = m(\omega_x^2x^2 + \omega_y^2y^2+\omega_z^2z^2)/2$, where $\omega_j$ represents the trapping frequencies and $m$ is the mass, which we take to be the same for both components.
We consider a cylindrically symmetric trap, $\omega_x=\omega_y\equiv\omega_\rho$, in the quasi-2D regime, $\omega_z \gg \omega_\rho$, where the $z$ direction is well described by the harmonic oscillator ground state. Scattering is still three-dimensional, i.e.~$a_{\alpha\beta}<<a_z$, where $a_{\alpha\beta}$ is the s-wave scattering length between components $\alpha$ and $\beta$, and $a_z = \sqrt{\hbar/ m\omega_z}$.
From here on we work with the planar coordinate, $\br = \{x,y\}$, and results will be given in terms of the radial length $a_\rho = \sqrt{\hbar/m\omega_\rho}$.

The solutions for each component $\psi_\alpha(\br)$ (where $\alpha=\{1,2\}$) are obtained by solving the coupled GPEs \cite{Esry1997,Pu1998,Ho1996},
\begin{equation}\label{Eq:GPE}
\left[H_{\rm sp}(\br) + \frac{\hbar^2}{m}\sum_\beta g_{\alpha\beta}n_\beta(\br)-\mu_\alpha \right]\psi_\alpha(\br) = \mathbf{0},
\end{equation}
where the condensate densities $n_\alpha = |\psi_\alpha(\br)|^2$ are individually unit-normalized, $\mu_\alpha$ are the corresponding chemical potentials, and the single-particle Hamiltonian is $H_{\rm sp}(\br) = -\hbar^2\nabla^2/2m + V(\br,z=0)$.
The dimensionless interactions are described by $g_{\alpha\beta}=N_\beta\sqrt{8\pi}a_{\alpha\beta}/a_z$, where $N_\beta$ is the number of atoms in component $\beta$.
In this work, we restrict our attention to balanced populations, $N_1=N_2\equiv N/2$, and intracomponent interactions, $g_{11}=g_{22}\equiv g$.
However, more general mixtures, along with a physical example, are discussed in Sec.~\ref{Sec:ExpConsid}.

\subsection{Variational Model}\label{Sec:Var}

In addition to numerically solving the GPEs, we derive analytical predictions for the stationary-state solutions, including a prediction for the miscible-to-immiscible phase transition.
To achieve this, motivated by the one-dimensional work of \cite{Navarro2009}, we implement a chirped gaussian ansatz (superscript $\mathcal V$) to approximate the wavefunctions,
\begin{equation}
\psi^{\mathcal V}_\alpha(\br)=A\exp\left[\frac{-(x\mp B)^2-y^2}{2W^2}\right] \exp\left[i(C \pm Dx+Ex^2)\right],
\end{equation}
where variational parameters describe the amplitude $A$, component separation $2B$, phase $C$, wavenumber $D$, chirp $E$ and width $W$. The phase is a product of the chemical potential and time, i.e.~$C=-\mu_\alpha^\mathcal{V} t/\hbar$.

As detailed in Appendix \ref{Sec:AppendixAna}, we use the Euler-Lagrange equations to derive equations of motion for the ansatz parameters.
We then go on to derive analytical stationary-state predictions for the density, $n_\alpha^\mathcal{V}=|\psi_\alpha^\mathcal{V}|^2$, in the miscible regime [Eqs.~(\ref{Eq:Amisc})-(\ref{Eq:muMisc})] and a set of transcendental equations for the immiscible regime [Eqs.~(\ref{Eq:cp0Imisc})-(\ref{Eq:muImisc})], both of which we will compare with our numerical results.
Finally, the variational prediction for the critical value of the intercomponent interaction strength, above which the system becomes immiscible, takes the form
\begin{equation}
g_{12}^{\rm c,\mathcal{V}} = g + 2\pi . \label{Eq:gc}
\end{equation}
The thermodynamic-limit result $ g_{12}^{\rm c,TD}=\sqrt{g_{11}g_{22}}=g$ is recovered for large particle number or intracomponent scattering length (recall that $g_{\alpha\beta}=N_\beta\sqrt{8\pi}a_{\alpha\beta}/a_z$).

\subsection{Two-Component Structure Factor}\label{Sec:SFformalism}

Bragg spectroscopy furnishes a means by which to selectively probe the $T=0$ component of the structure factor. Naturally, this provides a useful avenue for the study of quantum fluctuations by \textit{seeing through} the thermal fluctuations.
More specifically, Bragg spectroscopy measures the imaginary part of the response function and this relates to the dynamic structure factor (which will be defined shortly) as
\begin{equation}
\mathrm{Im}[\chi(\bk,\omega)] = -\frac{\pi}{\hbar}[S(\bk,\omega)-S(-\bk,-\omega)] ,
\end{equation}
where the finite-temperature contributions cancel to leading order \cite{Brunello2001,Blakie2002}.
Several groups have proposed schemes for the implementation of spin (and density) Bragg spectroscopy in binary condensates \cite{Rodriguez2002,Bruun2006,Carusotto2006,Baillie2016}. The general idea is to engineer a different coupling of the Bragg lasers to each component. This has been experimentally realized recently by the Vale group \cite{Hoinka2012}.

We find the excitations of the BEC by linearizing about the ground state using a small parameter $\eta$. This amounts to inserting the time-dependent ansatz
\begin{equation}
\Psi_\alpha(\br) = \left\{\psi_\alpha(\br) + \eta \left[ u_\alpha(\br)e^{-i\omega t} + v_\alpha^*(\br)e^{i\omega^*t} \right] \right\} e^{-i\mu_\alpha t/\hbar}
\end{equation}
into the GPE (Eq.~\ref{Eq:GPE})
and solving the resulting BdG equations.
Further details are provided in Appendix \ref{Sec:AppendixBog} (also see Refs.~\cite{Pu1998A,Timmermans1998,Ticknor2013,Ticknor2014}).

The $T=0$ component of the density $S_\mathcal{D}$ \cite{Zambelli2000} and spin $S_\mathcal{S}$ \cite{Abad2013,Symes2014A} structure factors are calculated from the BdG excitations, with each excitation labeled with superscript $\kappa$. Explicitly, we evaluate
\begin{align}
S_{\{\mathcal{D,S}\}} &(\bk,\omega) = \sum_\kappa \Big| \int d^2\br e^{i\bk\cdot\br} \Big([u^\kappa_1(\br)+v^\kappa_1(\br)]\psi_1(\br) \notag \\
& \pm [u^\kappa_2(\br)+v^\kappa_2(\br)]\psi_2(\br)\Big) \Big|^2 \delta(\omega-\omega^\kappa) , \label{Eq:SF}
\end{align}
where the $+~(-)$ indicates $S_\mathcal{D}$ ($S_\mathcal{S}$),  $\omega^\kappa = \epsilon^\kappa/\hbar$ (for BdG energy $\epsilon^\kappa$) and we have used the planar momentum coordinate $\bk = \{k_x,k_y\}$.  The corresponding static structure factors relate as $S_{\{\mathcal{D,S}\}}(\bk) = \int d\omega S_{\{\mathcal{D,S}\}}(\bk,\omega)$.

\begin{figure}
\begin{center}
\includegraphics[width=3.4in]{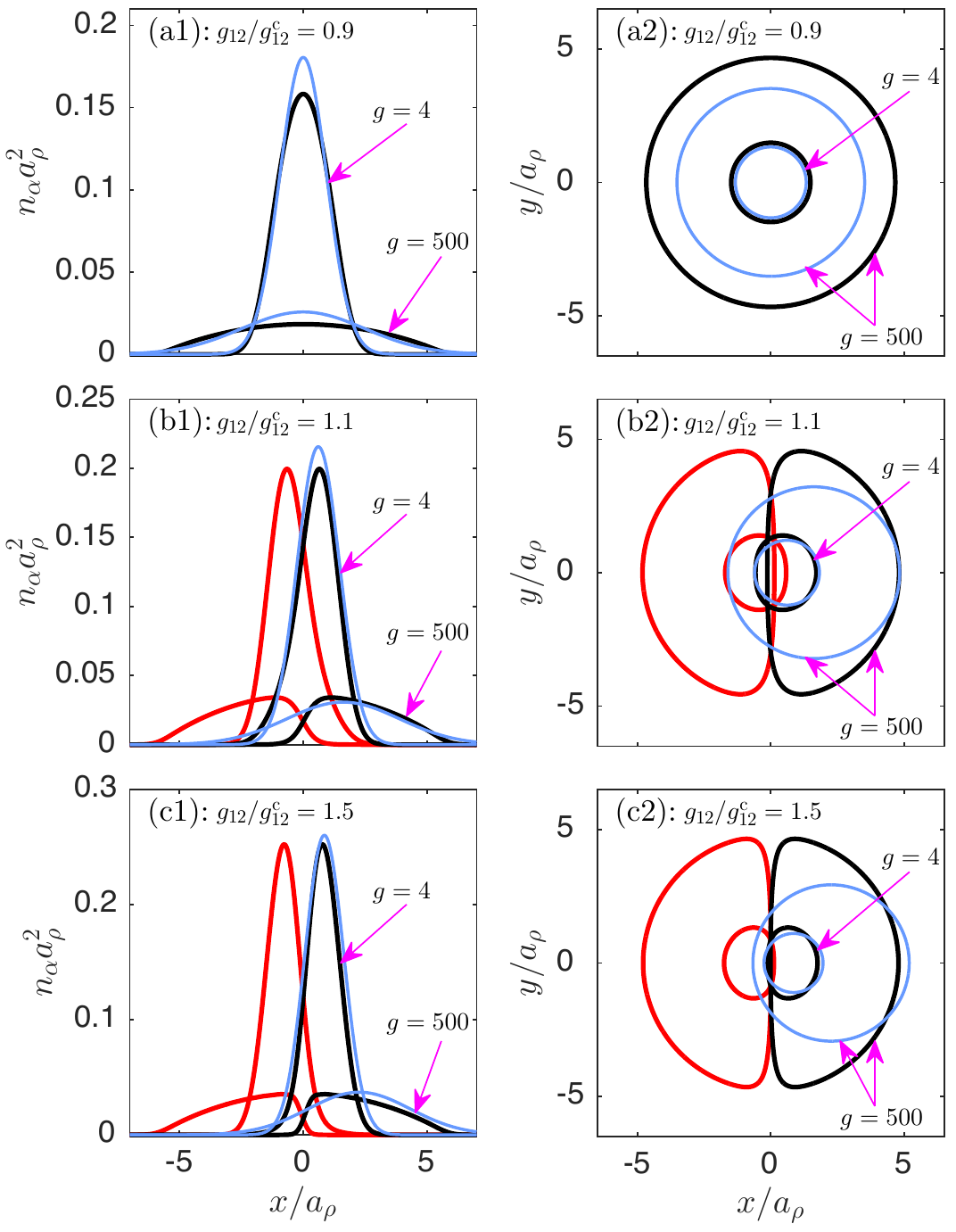}
\caption{Density profiles for (a) the miscible phase for $g_{12}/g_{12}^{\rm c}=0.9$, and the immiscible phase for (b) $g_{12}/g_{12}^{\rm c}=1.1$ and (c) $g_{12}/g_{12}^{\rm c}=1.5$, where $g_{12}^{\rm c}$ is the immiscibility critical point. The first component ($\alpha=1$) is shown in red (gray) while the second component ($\alpha=2$) is black. Column 1 offers a side view ($y=0$) while column 2 consists of top-down views, each exhibiting a single contour at $1/e$ of the peak density. The same two intracomponent interaction strengths are compared in all panels, i.e.~$g=\{4,500\}$.
The variational predictions $n_\alpha^\mathcal{V}$, shown as thin blue lines, are only plotted for the second component to avoid cluttering. The mixture is balanced, i.e.~$g_{11}=g_{22}\equiv g$ and $N_1=N_2$, where $g_{\alpha\beta}=N_\beta\sqrt{8\pi}a_{\alpha\beta}/a_z$, with $N_\beta$ being the number of atoms in component $\beta$, $a_{\alpha\beta}$ is the s-wave scattering length between components $\alpha$ and $\beta$, and $a_{\{\rho,z\}}=\sqrt{\hbar/m\omega_{\{\rho,z\}}}$.
\label{Fig:DenProfs}}
\end{center}
\end{figure}

\section{Results}

\subsection{The Immiscibility Phase Transition}\label{Sec:ResultsPT}

In this section we investigate how the immiscible critical value $g_{12}^{\rm c}$ shifts to higher values than the thermodynamic-limit prediction $g_{12}^{\rm c,TD}=g$. As will be shown in Sec.~\ref{Sec:ResultsSF}, when the intercomponent interaction strength lies within the resulting gap, i.e.~$g_{12}^{\rm c,TD}<g_{12}<g_{12}^{\rm c}$, then the quantum spin fluctuations become greatly enhanced.

In Fig.~\ref{Fig:DenProfs}, we compare our numerical and variational predictions for the density, using two intracomponent coupling strengths $g=\{4,500\}$ and three intercomponent couplings $g_{12}/g_{12}^{\rm c}=$ (a) 0.9, (b) 1.1 and (c) 1.5 \footnote{Note that for the numerical results we numerically determine $g_{12}^{\rm c}$, whereas for the variational results we take $g_{12}^{\rm c}=g_{12}^{{\rm c},\mathcal{V}}$ [Eq.~(\ref{Eq:gc})]}.
For the stronger coupling, $g=500$, the density is represented by the flatter, broader distribution in each subplot.
In the miscible phase [Fig.~\ref{Fig:DenProfs}(a)] the average density of both components is identical, while in the immiscible phase, Figs.~\ref{Fig:DenProfs}(b-c) show that the two components phase separate, forming a domain wall \cite{Trippenbach2000}, with a sharper separation for $g=500$.
As can be seen for $g=4$, the variational prediction agrees well with the full numerics for weak interactions, which is the regime of primary interest to this work.
For larger interactions, e.g.~$g=500$, the system enters the Thomas-Fermi regime where the Gaussian ansatz is less appropriate - this is most apparent in the top-down views in the immiscible phase [Figs.~\ref{Fig:DenProfs}(b2-c2)].
While in this regime a variational formulation based on Thomas-Fermi clouds may provide a better approximation, we will not pursue this avenue here.

\begin{figure}
\begin{center}
\includegraphics[width=3.4in]{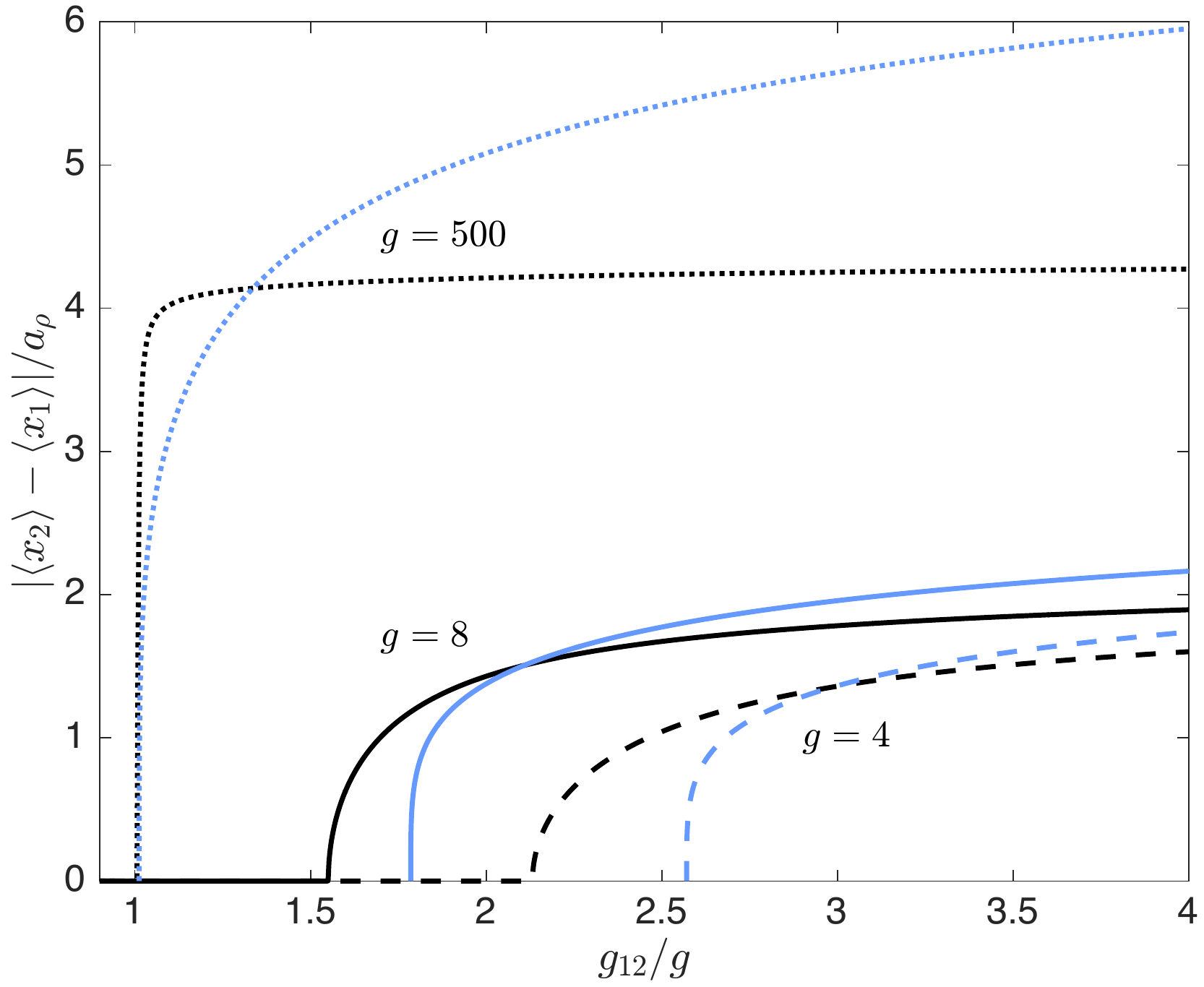}
\caption{Component separation (indicative of immiscibility) versus intercomponent coupling strength $g_{12}$ for various intracomponent couplings: $g = 4$ (dashed), $g=8$ (solid) and $g=500$ (dotted). Full numerical results are displayed in black while the variational results are in blue (gray). The position of component $\alpha$ is calculated according to $\langle x_\alpha\rangle=\int x \, n_\alpha(\br)d\br$, where we assume that the components always separate along the $x$ direction. For the variational model, $|\langle x_2\rangle-\langle x_1\rangle|=2B$.
 \label{Fig:Sepa}}
\end{center}
\end{figure}

To better characterize the onset of immiscibility, in Figure \ref{Fig:Sepa} we plot the component separation as a function of $g_{12}$.
Here it can be seen that the separation begins at a well-defined critical value $g_{12}^{\rm c}$ that increases with decreasing $g$. Overall, there is good qualitative agreement of the variational model (blue/gray) with the numerical results (black) for all cases.
Deep within the immiscible phase, $g_{12}\gg g_{12}^{\rm c}$, and for small $g$, the variational model does a particularly good job. However, as previously discussed, the quantitative agreement for the spatial condensate profiles deteriorates for large coupling ($g=500$) where the variational model is unable to capture the plateau seen for the numerical solution. This can be understood by noting from Figs.~\ref{Fig:DenProfs}(b-c) that the components are better able to avoid each other for the numerical solution, making it less sensitive to the precise value of $g_{12}$.

\begin{figure} 
\begin{center}
\includegraphics[width=3.3in]{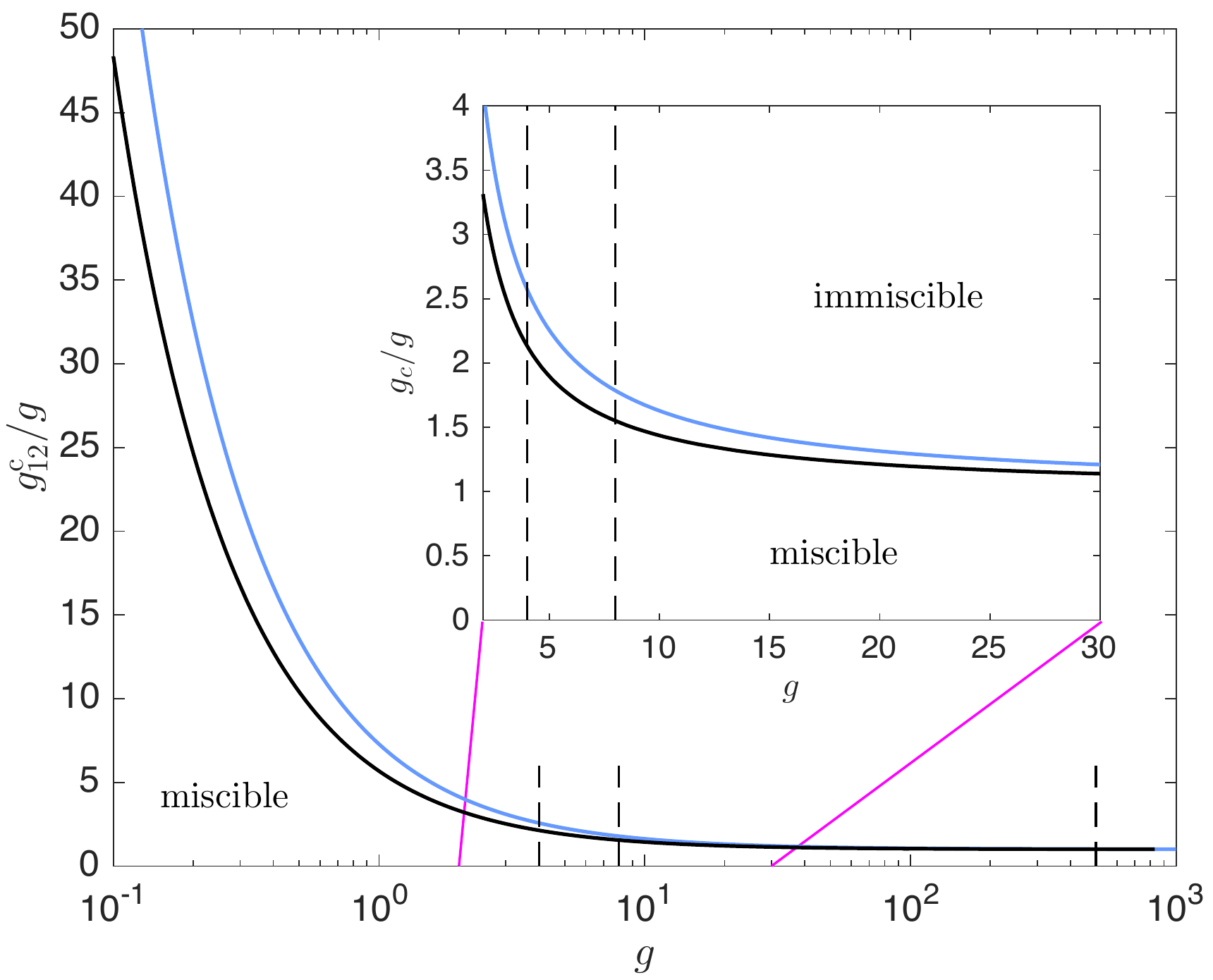}
\caption{Phase diagram for the immiscibility boundary. The critical intercomponent interaction strength $g_{12}^{\rm c}$ is plotted as a function of the intracomponent interaction strength $g$, where the miscible phase exists in the lower region. Full numerical results are in black, while the analytical prediction [Eq.~(\ref{Eq:gc})] is blue (gray). The vertical dashed lines represent the three interaction strengths that are the focus throughout the paper, i.e.~$g = \{4, 8, 500\}$. Inset: magnification of the region where the immiscibility boundary begins to significantly rise above the thermodynamic-limit result $g_{12}^{\rm c,TD}/g=1$.
 \label{Fig:PhaseD}}
\end{center}
\end{figure}

The immiscibility boundary is summarized by a phase diagram in Fig.~\ref{Fig:PhaseD}, with the numerical result in black and the variational one in blue (gray).
For large $g$, the boundary levels off to the thermodynamic-limit prediction $g_{12}^{\rm c,TD}/g=1$.
At the other extreme, where $g$ becomes small, the size of the miscible region grows as the role of quantum pressure increases and acts to stabilize the mixed phase.
The small-$g$ regime is precisely where the variational model is appropriate.
This explains why it captures the increasing trend of $g_{12}^{\rm c}/g$, yet the curve is qualitatively accurate throughout the region of parameters used.

\subsection{Enhanced Quantum Spin Fluctuations}\label{Sec:ResultsSF}

\begin{figure}
\begin{center}
\includegraphics[width=3.4in]{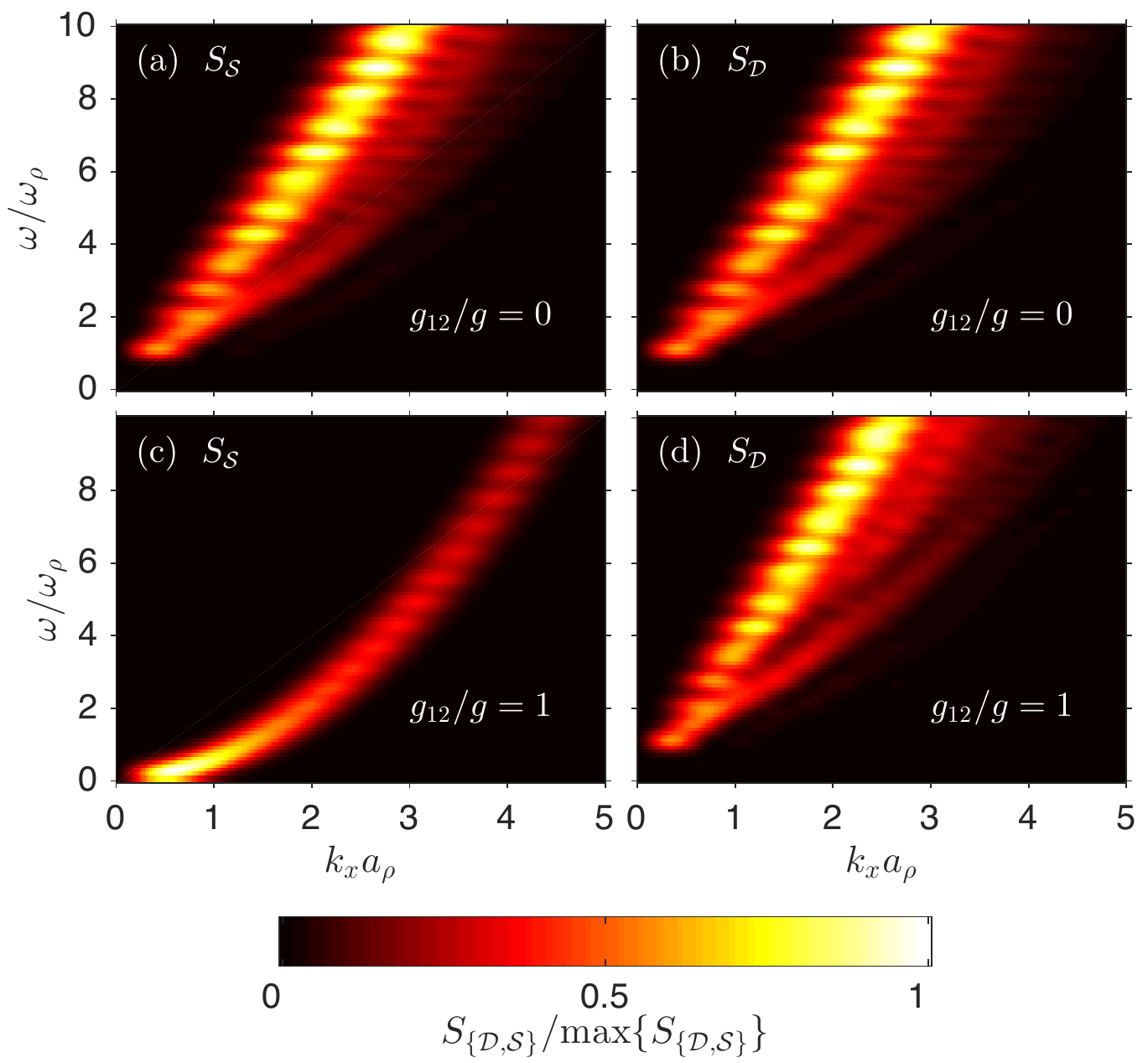}
\caption{Spin $S_{\mathcal{S}}(k_x,k_y=0,\omega)$ and density $S_{\mathcal{D}}(k_x,k_y=0,\omega)$ dynamic structure factor for the case of (a-b) uncoupled condensates ($g_{12}=0$), and (c-d) balanced intercomponent and intracomponent interactions ($g_{12}/g=1$), where $g = 500$.
 \label{Fig:DynSF}}
\end{center}
\end{figure}

\begin{figure*}
\begin{center}
\includegraphics[width=7in]{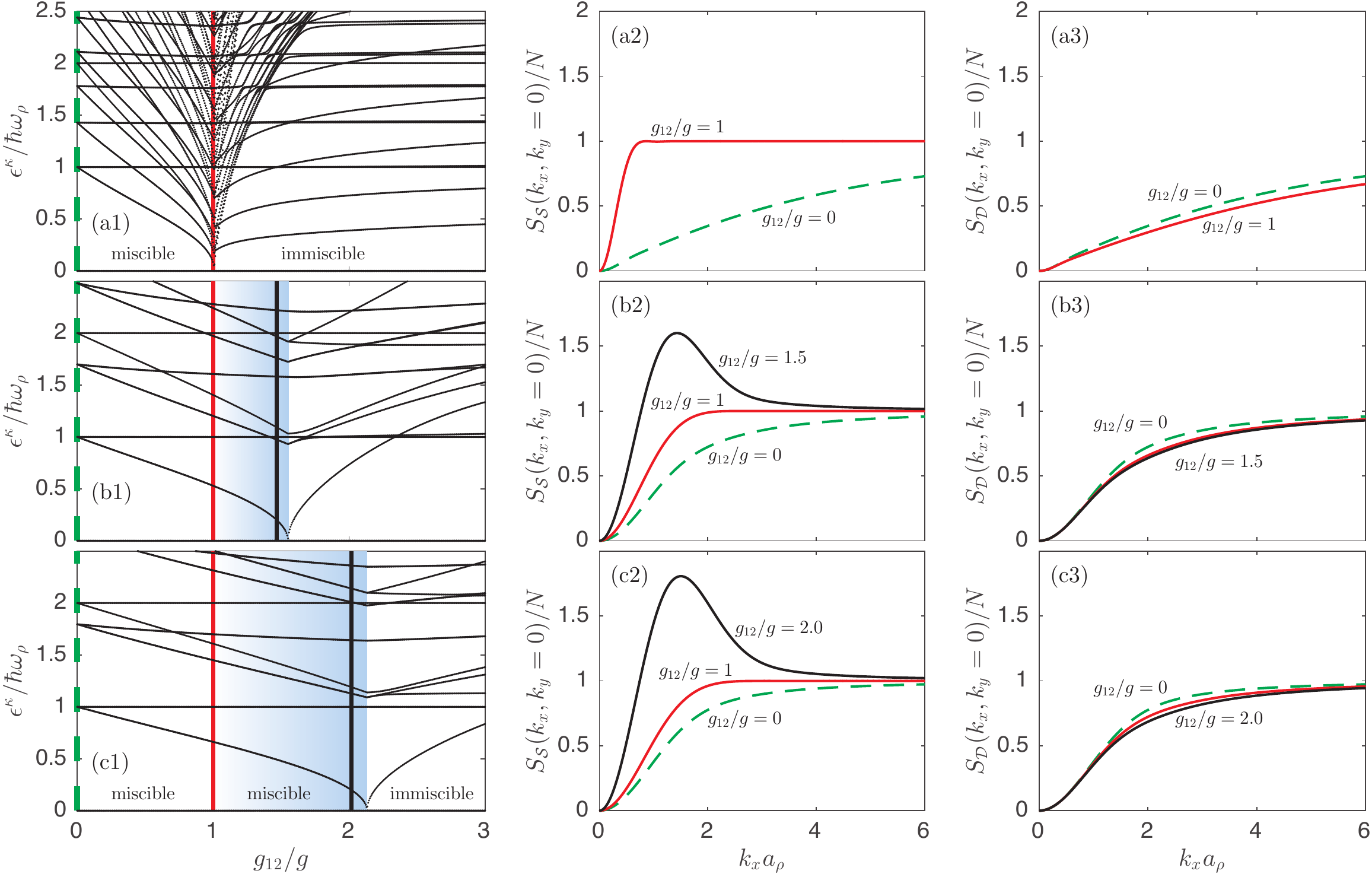}
\caption{(color online) Bogoliubov de Gennes energies versus inter-component interaction strength (column 1), and spin (column 2) and density (column 3) static structure factors versus momentum for the miscible phase. The nonlinearities considered are (a) $g=$ 500, (b) 8 and (c) 4. The vertical lines in column 1 correspond, with the same line styles, to the values of $g_{12}$ plotted in columns 2 and 3. The structure factors are normalized by the total condensate number $N=N_1+N_2$, where $N_1=N_2$. In the first column, the immiscibility transition occurs where an excitation energy softens to zero. The shading schematically indicates the region of enhanced quantum fluctuations.
 \label{Fig:SFandE}}
\end{center}
\end{figure*}

As we saw in Fig.~\ref{Fig:PhaseD}, the regime of small intracomponent interactions shifts the immiscibility phase transition higher than the thermodynamic-limit prediction, i.e.~$g_{12}^{\rm c}>g_{12}^{\rm c,TD}=g$.
This opens up a region of forced miscibility, $g<g_{12}<g_{12}^{\rm c}$, within which quantum spin fluctuations are greatly enhanced thanks to an effective attractive interaction \footnote{The interactions are effectively attractive in the sense that spin fluctuations act to decrease the interaction energy.}.
Bragg spectroscopy provides an especially useful tool for observing quantum fluctuations since it provides access to the $T=0$ component of the structure factor, even at finite $T$.
Thus, in this section, we numerically calculate the $T=0$ spin and density structure factors to show how the enhanced quantum fluctuations can be studied in experimentally accessible settings.

\subsubsection{Dynamic structure factor}

For reference, we first consider a BEC in the Thomas-Fermi regime, $g\gg1$, where the immiscibility transition occurs near the thermodynamic-limit prediction, $g_{12}^{\rm c}\approx g$, and the effective interactions experienced by the spin excitations are always repulsive.
When the two components are uncoupled, $g_{12}=0$, Figs.~\ref{Fig:DynSF} (a-b) show that the spin (left) and density (right), dynamic structure factors are identical, as expected.
On the other hand, a remarkable phenomenon emerges at the phase transition, where $g_{12}=g$. While here the density structure factor [Fig.~\ref{Fig:DynSF} (d)] is much the same as it was for $g_{12}=0$, apart from a slight hardening of the phonon excitations, the spin structure factor [Fig.~\ref{Fig:DynSF} (c)] is qualitatively different. In fact, the spin dispersion relation becomes parabolic.
This happens because the spin excitations become effectively noninteracting when $g_{12}=g$ since there is no preference for miscibility over immiscibility.

\subsubsection{BdG energies and the shifted phase transition}

We now extend the discussion to beyond the Thomas-Fermi regime to include small $g$, where spin excitations may now experience an effective attractive interaction.
The lowest BdG energies as a function of $g_{12}$ are plotted in the first column of Fig.~\ref{Fig:SFandE}, for $g=\{500,8,4\}$, from top to bottom.
The phase transition can be identified as where the energy of at least one excitation softens to zero. This occurs at the same point when approaching from either side, consistent with the second-order nature of the phase transition.
While in the Thomas-Fermi regime [Fig.~\ref{Fig:SFandE}(a1)] the transition occurs at $g_{12}^{\rm c}/g\approx1$, a considerable shift to $g_{12}^{\rm c}/g>1$ can be seen for small $g$ in Figs.~\ref{Fig:SFandE}(b1-c1).
The region of effective attractive interactions and enhanced quantum spin fluctuations, $g<g_{12}<g_{12}^{\rm c}$, is schematically shaded with blue (darker in grayscale) color.

\subsubsection{Static structure factor: enhanced quantum fluctuations}

Before proceeding, it is worth discussing the relationship between the structure factor and the sign of the interaction for a single-component, uniform system. Consider the Bijl-Feynman formula $S(\mathbf{k})=\epsilon_0(\mathbf{k})/\epsilon_B(\mathbf{k})$, where $\epsilon_0$ is the noninteracting dispersion relation and $\epsilon_B$ is the BdG energy \cite{Brunello2001}. If an excitation with some $\mathbf{k}$ experiences an effective attractive interaction, then the energy lies below its noninteracting counterpart, thus lifting $S(\mathbf{k})$ above unity as a result of atom bunching with respect to the noninteracting limit.

We present the numerical results for the spin (Fig.~\ref{Fig:SFandE}, column 2) and density (Fig.~\ref{Fig:SFandE}, column 3) static structure factors, restricting our attention to the miscible phase.
Consistent with the energy results of the first column, each row in Fig.~\ref{Fig:SFandE} considers one of $g=\{500,8,4\}$, from top to bottom.
The different curves are for various $g_{12}$ and the line styles are chosen to match (within each row) the values marked by vertical lines in column 1.
As a benchmark, consider the uncoupled case, $g_{12}=0$, indicated by the green dashed lines. Here, as expected, the spin $S_\mathcal{S}$ and density $S_\mathcal{D}$ structure factors are identical for a given $g$.
Increasing $g_{12}$ from zero has the opposite effect on $S_\mathcal{D}$ as it does on $S_\mathcal{S}$.
The density structure factor exhibits a reduction, indicating a further suppression of the density fluctuations and an increase of the overall effective interaction. The spin structure factor, on the other hand, markedly increases with increasing $g_{12}$, signaling an increase of the spin fluctuations and a reduction of the effective interaction.
In fact, once the intracomponent interaction has reached $g_{12}=g$ (red solid line), $S_\mathcal{S}$ becomes a flat line at unity for all three values of $g$. This indicates that the spin fluctuations are in the shot-noise regime and that the effective interactions have vanished (recall the Bijl-Feynman formula), as we already saw in Fig.~\ref{Fig:DynSF}(c) for $g_{12}=g$.
Note that the dip to zero at small momentum is due to the finite size of the system.
For $g=\{8,4\}$, a remarkable thing happens since we can have $g_{12}/g>1$ while still being in the miscible phase.
In this special regime [see shading in Figs.~\ref{Fig:SFandE}(b1-c1)] the effective interaction becomes attractive and spin fluctuations cause bunching with respect to the noninteracting limit, as indicated by the large peak of $S_\mathcal{S}$ in Figs.~\ref{Fig:SFandE} (b2-c2) (solid black lines). This constitutes an important result of the paper.

It is worth noting that even though quantum fluctuations are large, the condensate depletion need not be, unless of course the system is very close to the immiscibility transition.
This can be understood by the following. When interactions are effectively repulsive for a given excitation, the $v_\alpha^\kappa$ [see Eq.~(\ref{Eq:SF})] is approximately equal to $u_\alpha^\kappa$, but with opposite sign. The ensuing cancellation within Eq.~(\ref{Eq:SF}) results in a suppressed structure factor, implying anti-bunching with respect to the noninteracting limit. For attractive interactions, on the other hand, both $v_i$ and $u_i$ have the same sign, resulting in enhanced fluctuations (bunching). This sign reversal of $v_i$ is reminiscent of the weakly-interacting dipolar roton \cite{Santos2003}, for which the effective interaction is momentum dependent, switching from repulsive to attractive near the roton wavelength \cite{Bisset2013C,Blakie2013}.

\begin{figure}
\begin{center}
\includegraphics[width=3.4in]{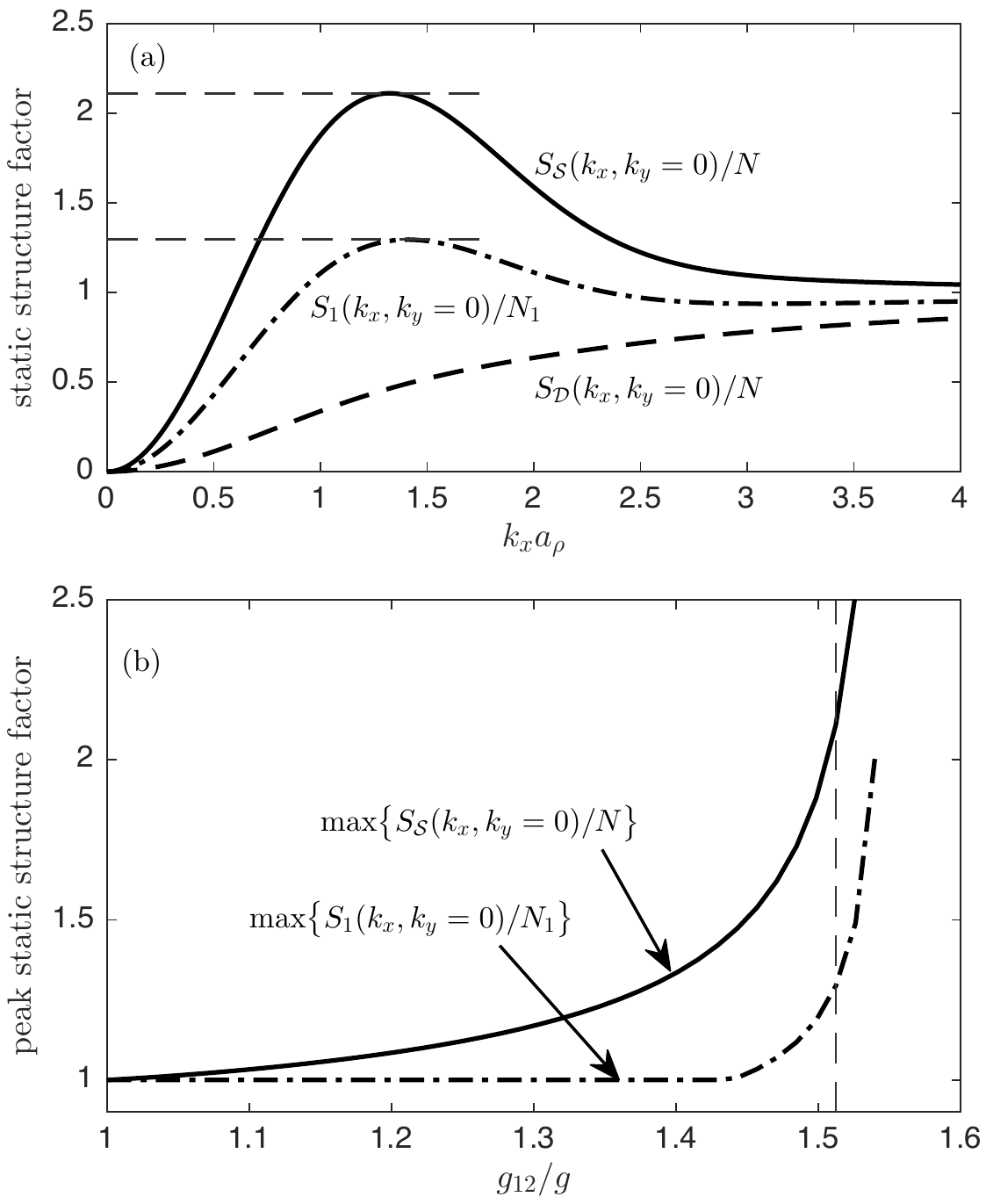}
\caption{(a) Density $S_\mathcal{D}$, spin $S_\mathcal{S}$, and one-component static structure factor $S_1$ for $g_{12}/g = 1.51$. The horizontal dashed lines indicate the structure factor peaks. (b) Spin and one-component static structure factor peaks as a function of $g_{12}/g$. The vertical dashed line indicates the interaction strength considered in (a). Note that $g_{12}^{\rm c}/g=1.55$ and $g=8$.
 \label{Fig:SFpeak}}
\end{center}
\end{figure}

\subsubsection{Single-component Bragg spectroscopy}

While spin Bragg spectroscopy should provide the most-direct signal of quantum spin fluctuations, this may not always be the easiest option in experiments. We now demonstrate that it is also possible to study quantum spin fluctuations with single-component Bragg spectroscopy of a binary condensate.
The single-component Bragg spectroscopy, corresponding to the single-component structure factor, can be calculated by restricting the integral of Eq.~(\ref{Eq:SF}) to the appropriate component, giving, for example,
\begin{align}
S_1 &(\bk,\omega) \\
 &= \sum_\kappa \Big| \int d^2\br e^{i\bk\cdot\br} [u^\kappa_1(\br)+v^\kappa_1(\br)]\psi_1(\br)
 \Big|^2 \delta(\omega-\omega^\kappa) \notag .
\end{align}
The one-component static structure factor then can be obtained as $S_1(\bk) = \int d\omega S_1(\bk,\omega)$.
The spin (solid), single-component (dot-dashed) and density (dashed) static structure factors are plotted in Fig.~\ref{Fig:SFpeak}(a) for $g_{12}/g=1.51$ and $g=8$. As expected, $S_1$ provides an intermediate measure between the spin and density structure factors; while the peak is somewhat subdued it is still clearly visible.
Figure \ref{Fig:SFpeak}(b) presents the spin (solid) and single-component (dot-dashed) structure factor peaks [indicated by horizontal lines in Fig.~\ref{Fig:SFpeak}(a)] as a function of $g_{12}$. Here it can be seen that close to the immiscibility phase transition, which is at $g_{12}^{\rm c}=1.55$, the spin contribution dominates and $S_1$ provides an effective measure of the enhanced quantum spin fluctuations.

\section{Experimental considerations}\label{Sec:ExpConsid}

From a qualitative perspective we expect our results to be quite general. Binary condensates can be constructed from a wide variety of mixtures, such as by combining different elements, isotopes, hyperfine states, or even spin states. To realize enhanced quantum spin fluctuations, the key ingredients are for both components to be miscible while, at the same time, $g_{12}>\sqrt{g_{11}g_{22}}$. This is possible if the immiscibility critical point is shifted higher than the thermodynamic-limit prediction $g_{12}^{\rm c,TD} = \sqrt{g_{11}g_{22}}$ (recall Fig.~\ref{Fig:PhaseD}).
Our proposal relies on achieving this by enhancing the role of quantum pressure.

To give a flavor for the different kinds of experimental regimes that could realize enhanced quantum spin fluctuations, consider $^{87}$Rb, for which a mixture can be constructed from the hyperfine states $|F,m_F\rangle=|2,-1\rangle$ and $|1,1\rangle$. For these, there is a Feshbach resonance that can control the interspecies scattering length \cite{Nicklas2015,Widera2004,Erhard2004,Gross2010}. Let us then take $g=8$ as a representative example for the results in this paper. Since the intraspecies scattering lengths are $a_{11}=a_{22}=100a_0$, then choosing a trap with frequencies $\{\omega_\rho,\omega_z\}=2\pi\times\{1,10\}$ Hz would correspond to $N_\alpha=1030$ atoms per component.

To increase the atom number, one can consider species with smaller scattering lengths and/or masses. Furthermore, it is not necessary for both intraspecies scattering lengths, or even masses, to be equal. A possible complication for the imbalanced case, though, is that inhomogeneous trapping potentials can cause the binary mixture to become immiscible even when $g_{12}<\sqrt{g_{11}g_{22}}$ \cite{Lee2016}, hence shifting $g_{12}^{\rm c}$ in the wrong direction. Fortunately, this negative $g_{12}^{\rm c}$ shift can be reversed by imbalancing the atom numbers in favor of the component with the strongest interactions. In any case, the enhanced quantum pressure in the regime of interest will act to raise $g_{12}^{\rm c}$ as required.

Although the flattened trap geometry considered in this paper has some advantages, such as for {\it in situ} imaging of the fluctuations and numerical tractability, our results should be qualitatively extendible to more spherical geometries. In fact, loosening the trap along the tight direction should allow the regime of enhanced quantum fluctuations to extend to larger atom numbers.

Not only are quantum spin fluctuations enhanced on approach to the transition, but so too are thermal fluctuations \cite{Bisset2015}.
Fortunately, though, Bragg spectroscopy provides a window through which one can selectively observe the quantum contribution, even at finite $T$.
To appreciate this, we estimate the temperature that would otherwise be required for quantum fluctuations to dominate. The excitations primarily responsible for enhanced fluctuations have energies that soften on approach to the transition and, as can be seen in Figs.~\ref{Fig:SFandE}(b1) and \ref{Fig:SFandE}(c1), these typically have energies of the same order as the radial confinement energy $\hbar\omega_\rho$. Again choosing $\omega_\rho=2\pi\times 1$ Hz means that without Bragg spectroscopy, for quantum fluctuations to dominate, experiments would otherwise need to be in the very-low-temperature regime, $T< 0.1$ nK.
Concerning the validity of the quasi-2D approximation for the candidate regime above, we note that for $g=8$ the chemical potential relative to that of the noninteracting ground state is $\mu/\hbar\omega_z = 0.149$, where $\mu_1=\mu_2\equiv\mu$, thus satisfying the requirement that $\mu/\hbar\omega_z \ll 1$.

\section{Conclusions}\label{Sec:Conclusions}

In quantum fluids, quantum fluctuations are typically not important unless the dimensionality is low, the density is high or the particle number is very low. We have proposed an alternative scheme for enhanced quantum fluctuations that does not require any of these conditions to be satisfied, but instead relies on engineering an effective attractive interaction for the spin excitations of a two-component BEC. A strength of this approach is that the BEC is not vulnerable to dynamic instability, as would be the case for true attractive interactions.

In addition to numerical calculations, we developed a variational model to corroborate the phase diagram of the shifted immiscibility phase transition. Such a shift of $g_{12}^{\rm c}$ is crucial for reaching the regime of enhanced quantum fluctuations.
We find excellent qualitative and good quantitative agreement between the numerical findings and the variational model, and the latter should be useful when scouting for experimentally relevant regimes.
We numerically performed BdG calculations to show that the enhanced quantum spin fluctuations should be observable using either spin or single-component Bragg spectroscopy for two-component BECs.

This work opens a number of possibilities for future research. An obvious direction is for experiments to quantitatively test beyond-mean-field theories. Since the size of the quantum fluctuations is tunable, by adjusting $g_{12}$ to approach the immiscibility transition, it would be interesting to see at which point, for example, BdG theory breaks down.
It is also an open question as to how quantum fluctuations might shift the immiscibility transition.
Moreover, the proposed regime could be useful for quantitative comparisons between beyond-mean-field theories with Monte Carlo simulations.
Another intriguing direction would be to go beyond BdG theory to study the quantum critical physics on approach to the immiscibility phase transition.

As a final note, the enhanced fluctuations should also be directly observable via {\it in situ} imaging of the spin density ($n_1-n_2$). Such a scheme was theoretically investigated for thermal spin fluctuations in Ref.~\cite{Bisset2015}. However, in this case the quantum and thermal fluctuations will both contribute, and the temperature will need to be quite low to suppress the thermal contribution.\\

\noindent {\bf Acknowledgments: $\,$}
We thank D.~Baillie, T.~Bienaim\'e, P.~B.~Blakie, R.~Carretero-Gonz{\'a}lez, G. Ferrari, S.~I.~Mistakidis, A.~Recati and C.~Qu for useful and stimulating discussions.
RNB acknowledges the EU QUIC project and Provincia Autonoma di Trento for financial support.
PGK acknowledges the support of NSF-DMS-1312856 and PHY-1602994, 
from the ERC under FP7, Marie Curie Actions, People, International Research Staff Exchange Scheme (IRSES-605096), and the Stavros Niarchos Foundation via the Greek Diaspora Fellowship Program.
CT acknowledges support by Los Alamos National Laboratory, which is operated by LANS, LLC for the NNSA of the U.S. DOE and, specifically, Contract No. DEAC52-06NA25396.

\appendix

\section{Two-component Gaussian ansatz}\label{Sec:AppendixAna}

To gain a theoretical handle on the miscibility transition in the quasi-2D setting we apply a chirped Gaussian ansatz (superscript $\mathcal V$) as an approximation to the wavefunctions (similar to what was done in 1D by \cite{Navarro2009}), i.e.,
\begin{equation}
\psi^{\mathcal V}_\alpha(\br)=A\exp\left[\frac{-(x\mp B)^2-y^2}{2W^2}\right] \exp\left[i(C \pm Dx+Ex^2)\right],
\end{equation}
where $A$ is the amplitude, $B$ is the component separation and $W$ determines the width. The phase $C$ is is related to the chemical potential and time according to $C=-\mu_\alpha^\mathcal{V} t/\hbar$, while $D$ and $E$ represent the wavenumber and chirp, respectively. The upper of $\mp$ and $\pm$ are for component $\alpha=1$ while the lower are for $\alpha=2$.

We utilize the Euler-Lagrange equations,
\begin{equation}
\frac{\partial \mathcal L}{\partial \mathcal{P}_j} - \frac{d}{dt}\left(\frac{\partial \mathcal L}{\partial \dot{\mathcal{P}_j}} \right) = 0 ,
\end{equation}
with $\mathcal{P}_j$ = \{$A$, $B$, $C$, $D$, $E$, $W$\}. The Lagrangian takes the form
\begin{equation}
\mathcal{L} = \int_{-\infty}^\infty ( L_1 + L_2 + L_{12} ) d^2\br ,
\end{equation}
where
\begin{align}
L_\alpha &= \frac{\hbar^2}{2m} \left( \left|\frac{\partial \psi_\alpha^{\mathcal{V}}}{\partial x}\right|^2 + \left|\frac{\partial \psi_\alpha^{\mathcal{V}}}{\partial y}\right|^2 \right) \\
&+ \frac{m\omega_\rho^2}{2}\left(x^2+y^2 \right)\left|\psi_\alpha^{\mathcal{V}}\right|^2 + \frac{\hbar^2g}{2m} \left|\psi_\alpha^{\mathcal{V}}\right|^4 \\
&+ \frac{i\hbar}{2} \left(\psi_\alpha^\mathcal{V}\frac{\partial{\psi_\alpha^\mathcal{V}}^*}{\partial t} - {\psi^\mathcal{V}_\alpha}^*\frac{\partial\psi_\alpha^\mathcal{V}}{\partial t}\right) ,
\end{align}
and the intercomponent coupling is given by
\begin{equation}
L_{12} = \frac{\hbar^2g_{12}}{m}\left|\psi_1^\mathcal{V}\right|^2\left|\psi_2^\mathcal{V}\right|^2.
\end{equation}
Recall that we have taken the symmetric situation $g_{11}=g_{22}\equiv g$ (and also $\mu_1=\mu_2\equiv \mu$), where $g_{\alpha\beta}=N_\beta\sqrt{8\pi}a_{\alpha\beta}/a_z$ with $N_\beta$ being the number of atoms in component $\beta$, $a_{ij}$ is the scattering length between component $\alpha$ and $\beta$, and $a_z = \sqrt{\hbar/m\omega_z}$ is the confinement length in the tight direction.

\subsection*{Equations of motion}

For clarity, we now switch to dimensionless form: $\tilde\mu = \mu/\hbar\omega_
\rho$; $\tilde t = t\omega_\rho$; $\tilde{A}=a_\rho A$; $\tilde{B}=B/a_\rho$; $\tilde{C}=C$; $\tilde{D}=a_\rho D$; $\tilde{E}=a_\rho^2 E$; $\tilde{W}=W/a_\rho$, where $a_\rho = \sqrt{\hbar/m\omega_\rho}$.
Simultaneously solving the Euler-Lagrange equations we obtain the equations of motion for the ansatz parameters:
\begin{equation}
\frac{d\tilde A}{d\tilde t} = -2 \tilde A \tilde E , \label{Eq:dA_dt}
\end{equation}
\begin{equation}
\frac{d\tilde B}{d\tilde t} = \tilde D + 2\tilde B \tilde E ,
\end{equation}
\begin{align}
\frac{d\tilde C}{d\tilde t} &= \frac{\tilde B^2}{\tilde W^4}  - \frac{\tilde D^2}{2} -\frac{1}{\tilde W^2}- \frac{\tilde B^2}{2} -\frac{3\tilde A^2g}{4} + \frac{\tilde A^2 \tilde B^2 g}{2\tilde W^2} \notag \\
 & - \tilde A^2e^{-2\tilde B^2/\tilde W^2}g_{12} \left(\frac{3}{4} + \frac{\tilde B^4}{\tilde W^4} \right) , \label{Eq:dC_dt}
\end{align}
\begin{equation}
\frac{d\tilde D}{d\tilde t} = \tilde B - 2 \tilde D \tilde E - \frac{2\tilde B}{\tilde W^4} -\frac{\tilde A^2\tilde Bg}{\tilde W^2} + \frac{2\tilde A^2\tilde B^3g_{12}}{\tilde W^4}e^{-2\tilde B^2/\tilde W^2} ,
\end{equation}
\begin{align}
\frac{d\tilde E}{d\tilde t} &= \frac{1}{\tilde W^4} - 1 - 2\tilde E^2 + \frac{\tilde A^2g}{2\tilde W^2} \notag\\
& + \frac{\tilde A^2g_{12}}{2\tilde W^4}e^{-2\tilde B^2/\tilde W^2}(\tilde W^2-2\tilde B^2) ,
\end{align}
\begin{equation}
\frac{d\tilde W}{d\tilde t} = 2\tilde E\tilde W . \label{Eq:dW_dt}
\end{equation}

\subsection*{Stationary state solutions for the miscible phase}

We look for stationary-state solutions by setting the left hand side of the equations of motion [(\ref{Eq:dA_dt})-(\ref{Eq:dW_dt})] to zero, with the exception of Eq.~(\ref{Eq:dC_dt}) for which the phase continues to evolve steadily according to the chemical potential, $d\tilde C/d\tilde t = -\tilde\mu$. For such a state, one can immediately see that the wavenumber and chirp are zero, i.e.~$D_0=E_0=0$.

By definition, the miscible phase has $B_0=0$, and solving the resulting equations produces analytic expressions for the equilibrium amplitude and width, respectively:
\begin{align}
\tilde A_0^2 &= \frac{4}{3(g+g_{12})} \left(2\tilde\mu_0 - \sqrt{\tilde\mu^2_0+3} \right) ,\label{Eq:Amisc} \\
\tilde W_0^2 &= \frac{1}{3}\left(\tilde\mu_0 + \sqrt{\tilde\mu^2_0+3} \right),
\end{align}
where the chemical potential is
\begin{equation}
\tilde\mu_0 = \frac{1}{\sqrt{8\pi}} \frac{(3g + 3g_{12} + 4\pi)}{\sqrt{g+g_{12} + 2\pi}} . \label{Eq:muMisc}
\end{equation}

\subsection*{Stationary state solutions for the immiscible phase}

Finding stationary state solutions for the immiscible phase ($B_0\neq0$) is not so straightforward and the equations of motion, (\ref{Eq:dA_dt})-(\ref{Eq:dW_dt}), reduce to a set of transcendental equations that we solve numerically:

\begin{align}
0 &=  \frac{1}{\tilde W^2_0} + \frac{3\tilde A^2_0g}{4} + \frac{\tilde B^2_0}{2} -\frac{\tilde B^2_0}{\tilde W^4_0} - \frac{\tilde A^2_0\tilde B^2_0g}{2\tilde W^2_0} \notag \\
& + \tilde A^2_0g_{12}e^{-2\tilde B^2_0/\tilde W^2_0} \left(\frac{3}{4} + \frac{\tilde B^4_0}{\tilde W^4_0} \right) - \tilde\mu_0 ~, \label{Eq:cp0Imisc}
\end{align}
\begin{equation}
0 = \tilde B_0 - \frac{2\tilde B_0}{\tilde W^4_0} - \frac{\tilde A^2_0\tilde B_0g}{\tilde W^2_0} + \frac{2\tilde A^2_0 \tilde B^3_0 g_{12}}{\tilde W^4_0} e^{-2\tilde B^2_0/\tilde W^2_0} ,
\end{equation}
\begin{equation}
0 = \frac{1}{\tilde W^4_0} + \frac{\tilde A^2_0g}{2\tilde W^2_0} - 1 + \frac{\tilde A^2_0g_{12}}{2\tilde W^4_0}e^{-2\tilde B^2_0/\tilde W^2_0} \left(\tilde W^2_0 - 2\tilde B^2_0 \right) ,
\end{equation}
where the chemical potential is given by
\begin{equation}
\tilde\mu_0 = \frac{\pi \tilde A^2_0\tilde W^2_0}{2} \left(\frac{1}{\tilde W^2_0} + \tilde A^2_0g + \tilde B^2_0 + \tilde W^2_0 + \tilde A^2_0 g_{12}e^{-2\tilde B^2_0/\tilde W^2_0} \right). \label{Eq:muImisc}
\end{equation}

\subsection*{Critical interaction strength}

To find the transition point between miscible and immiscible phases we construct the Jacobian matrix, $\mathcal J_{jk} = \partial \mathcal F_j/\partial \mathcal P_k$, where $\mathcal F_j=d\mathcal P_j/dt$ are the equations of motion given (in dimensionless form) by Eqs.~(\ref{Eq:dA_dt})-(\ref{Eq:dW_dt}).
The stationary-state solution for the miscible phase, Eqs.~(\ref{Eq:Amisc})-(\ref{Eq:muMisc}), is substituted into the Jacobian matrix and we subsequently calculate its eigenvalues.
Finding the interaction strengths at which the appropriate eigenvalue vanishes then gives an analytic prediction for the critical value of $g_{12}$, i.e.,
\begin{equation}
g_{12}^{c,\mathcal{V}} = g + 2\pi .
\end{equation}

\section{Two-component Bogoliubov-de Gennes theory}\label{Sec:AppendixBog}

For continuity of notation, we continue to work in the quasi-2D regime with planar coordinates, $\br = \{x,y\}$, although the generalization to 3D is straightforward.
Recall, that in the weakly interacting limit, the condensate components $\psi_\alpha(\br)$ (with $\alpha = \{1,2\}$) are computed by solving the coupled GPEs,
\begin{equation}
H_\alpha^{GP}\psi_\alpha(\br) \equiv \left[H_{\rm sp}(\br) +C_\alpha(\br)-\mu_\alpha \right]\psi_\alpha(\br) = \mathbf{0},
\end{equation}
with the single particle Hamiltonian $H_{\rm sp}(\br) = -\hbar^2\nabla^2/2m + V(\br)$, where $V(\br)$ is the trap potential.
The interactions are described by
\begin{align}
C_\alpha(\br) &= \sum_\beta \int d^2 \br^\prime U_{\alpha\beta}(\br-\brp) n_\beta(\br^\prime) ~, \\
&= \frac{\hbar^2}{m}\sum_\beta g_{\alpha\beta}n_\beta(\br) ~,
\end{align}
where the condensate density $n_\alpha$ is unit-normalized $\int  n_\alpha(\br) d^2\br = \int |\psi_\alpha(\br)|^2 d^2\br = 1$ 
 and we consider the case of contact interactions, i.e.~$U_{\alpha\beta}(\br-\brp)=\delta(\br-\brp)\hbar^2g_{\alpha\beta}/m$ with $g_{\alpha\beta}=N_\beta\sqrt{8\pi}a_{\alpha\beta}/a_z$ and $N_\beta$ is the number of atoms in condensate $\beta$.

Excitations are found by linearizing about the ground state. Using the ansatz,
\begin{equation}
\Psi_\alpha(\br) = \left\{\psi_\alpha(\br) + \eta \left[ u_\alpha(\br)e^{-i\omega t} - v_\alpha^*(\br)e^{i\omega^*t} \right] \right\} e^{-i\mu_\alpha t/\hbar} ,
\end{equation}
and keeping terms up to first-order in the small parameter $\eta$, results in the two-component Bogoliubov de Gennes equations \cite{Pu1998A,Timmermans1998,Ticknor2013},
\begin{align}
\left( \begin{array}{cc}
H^{GP}_\alpha & 0 \\
0 & -H^{GP}_\alpha
\end{array} \right)
\left( \begin{array}{c}
u_{\alpha}^{\kappa} \\ v_{\alpha}^{\kappa}
\end{array} \right)
&+ \sum_{\beta=1,2}\left( \begin{array}{cc}
X_{\alpha\beta}^a & -X_{\alpha\beta}^b \\
{X_{\alpha\beta}^b}^* & -X_{\alpha\beta}^a
\end{array} \right)
\left( \begin{array}{c}
u_{\beta}^{\kappa} \\ v_{\beta}^{\kappa}
\end{array} \right) \notag \\ 
&= \epsilon^\kappa \label{Eq:BdG}
\left( \begin{array}{c}
u_{\alpha}^{\kappa} \\ v_{\alpha}^{\kappa}
\end{array} \right) ,
\end{align}
where the superscript index $\kappa$ labels each excitation.
The condensate exchange operators act on a test function $f$ according to
\begin{align}
(X_{\alpha\beta}^af)(\br) &= \psi_\alpha(\br)\int d^2\brp U_{\alpha\beta}(\br-\brp)\psi_\beta^*(\brp)f(\brp) , \notag \\
&= \frac{\hbar^2}{m}\psi_\alpha(\br)\psi_\beta^*(\br)f(\br)g_{\alpha\beta} , \\
(X_{\alpha\beta}^bf)(\br) &= \psi_\alpha(\br)\int d^2\brp U_{\alpha\beta}(\br-\brp)\psi_\beta(\brp)f(\brp) , \notag \\
&= \frac{\hbar^2}{m}\psi_\alpha(\br)\psi_\beta(\br)f(\br)g_{\alpha\beta} , \\
({X_{\alpha\beta}^b}^*f)(\br) &= \psi_\alpha^*(\br)\int d^2\brp U_{\alpha\beta}(\br-\brp)\psi_\beta^*(\brp)f(\brp) , \notag \\
&= \frac{\hbar^2}{m}\psi_\alpha^*(\br)\psi_\beta^*(\br)f(\br)g_{\alpha\beta} .
\end{align}
Finally, quasiparticle modes are normalized according to
\begin{equation}
\sum_{\alpha=1,2}\int d^2\br \Big(|u_\alpha^\kappa(\br)|^2-|v_\alpha^\kappa(\br)|^2 \Big) = 1 .
\end{equation}

We note, that it is important to ensure that all excitations are orthogonal to the ground state \cite{Morgan2000}; to achieve this, we apply a condensate-projector before and after the exchange operators.
We solve Eqs.~\ref{Eq:BdG} using a spectral basis of non-interacting harmonic oscillator modes, and utilize the Gauss-Hermite quadrature.

\bibliographystyle{apsrev4-1}


%

\end{document}